\documentclass{sig-alternate-05-2015}
\usepackage{amsmath}
\usepackage{booktabs}
\usepackage{graphicx}
\usepackage{textcomp}
\usepackage[linesnumbered,boxed]{algorithm2e}
\usepackage{pifont}
\usepackage{caption}
\usepackage{dblfloatfix}
\usepackage{footnote}
\renewcommand\footnoterule{{\hrule height 1pt}}
\newsavebox\mybox
\usepackage[labelfont=bf]{caption}

\SetKwInput{KwInput}{Input}
\SetKwInput{KwOutput}{Output}
\usepackage{caption}
\usepackage[labelfont=it,textfont=scriptsize,justification=justified]{subcaption}

\begin{document}

\setcopyright{acmcopyright}

\doi{http://dx.doi.org/xx.xxxx/xxxxxxx.xxxxxxx}

\isbn{978-1-4503-4486-9/17/04}


\acmPrice{\$15.00}

%
\conferenceinfo{SAC'17,}{ April 3-7, 2017, Marrakesh, Morocco}
\CopyrightYear{2017} 

\title{APR: Architectural Pattern Recommender \\ 
}
%
%
%
%
%

\numberofauthors{2} 
%
\author{
%
%
\alignauthor
Shipra Sharma\\
       \affaddr{Indian Institute of Technology Ropar}\\
       \affaddr{Rupnagar, Punjab}\\
       \affaddr{India}\\
       \email{shipra.sharma@iitrpr.ac.in}
\alignauthor
Balwinder Sodhi\\
       \affaddr{Indian Institute of Technology Ropar}\\
       \affaddr{Rupnagar, Punjab}\\
       \affaddr{India}\\
       \email{sodhi@iitrpr.ac.in}
}
\CopyrightYear{2017}
\setcopyright{othergov}
\conferenceinfo{SAC 2017,}{April 03-07, 2017, Marrakech, Morocco}
\isbn{978-1-4503-4486-9/17/04}\acmPrice{\$15.00}
\doi{http://dx.doi.org/10.1145/3019612.3019780}
\maketitle
\begin{abstract}
This paper proposes Architectural Pattern Recommender (APR) system which helps in such architecture selection process. Main contribution of this work is in replacing the manual effort required to identify and analyse relevant architectural patterns in context of a particular set of software requirements. Key input to APR is a set of architecturally significant use cases concerning the application being developed. Central idea of APR's design is two folds: a) transform the unstructured information about software architecture design into a structured form which is suitable for recognizing textual entailment between a requirement scenario and a potential architectural pattern. b) leverage the rich experiential knowledge embedded in discussions on professional developer support forums such as Stackoverflow to check the sentiment about a design decision. APR makes use of both the above elements to identify a suitable architectural pattern and assess its suitability for a given set of requirements. Efficacy of APR has been evaluated by comparing its recommendations for ``ground truth'' scenarios (comprising of applications whose architecture is well known). 

\end{abstract}



\section{Introduction}


Architecture of a software essentially consists of basic building blocks of the software system. Often it is possible to arrange these blocks in more than one manner to achieve a specific functionality. An \emph{architectural pattern} ascribes to recurring software development problem scenarios and how they can be solved by following a certain design approach. The architect uses existing architectural patterns to select an optimal arrangement of system's building blocks. This process, being a creative one, is often performed manually by the architects and ``quality'' of its outcomes depends largely on his/her experience and skills. 

In practice, the choice of an architecture is influenced by the concerns of various stakeholders, and thus is a complex task. Because designing architecture of a software application is a creative activity, there is no fixed recipe to select an architectural pattern that can be used as a blueprint of a system's architecture. As Salah et. al.\cite{Salah2013} have noted, superficial or limited knowledge of patterns may lead to wrong architectural design thereby adversely affecting implementation and maintenance of the complete system. In other words, extensive and experiential knowledge play a significant role in this domain. To identify patterns that can suit a given scenario, usually, a software architect goes through the standard text or relies on organization's memory of similar past work. Next, from among the patterns thus selected, those which have been earlier successful in a similar scenario are shortlisted. Finally, based on his/her judgment, the architect selects one of the patterns. We propose a recommender system, henceforth called APR, to semi-automate the process of architectural pattern selection and suggest relevant patterns according to the software requirements. APR's objective is to replace the manual effort required in this process by recommending suitable architectural patterns to work with. To achieve this we use semantic and contextual information retrieval techniques. It has been observed that less experienced software architects and developers often avoid using patterns due to the complexity in selecting an appropriate one\cite{Salah2013} -- APR is expected to benefit such users the most.


One of the key inputs needed to design the architecture of a system is software's requirements specifications. Some requirements can also serve as design decisions\cite{van2008software}. Thus, in our approach we use system requirements as an input to the APR. Currently the APR focuses only on recommending one pattern at a time\footnote{Normally, a complex software system employs more than one architectural pattern.}. The paper is structured as follows. Section-\ref{relwork} gives an insight into the related recommenders and techniques. The overall design and our approach is explained in detail in Section-\ref{mymethod}. Experimental findings and analysis are discussed in Section-\ref{Analy} along with validity of our results. Finally, Section-\ref{confut} concludes with description of our future direction.

\section{Related Work}\label{relwork}
Selection of architectural pattern can be done in various ways. A common approach which is also most time consuming is the manual approach where one would study the architecture related literature to identify candidate architecture for the problem at hand. Another approach is based on using an indexed knowledge-base where keyword based search is used to retrieve and identify relevant architecture literature/information. Systems such as \cite{Sanyawong2014} and \cite{Salah2013} are the examples of such kinds. One of the limitations of these techniques is that relying purely on textual similarity when searching the knowledge-base often leads to results which are less accurate than manually sifting through the information. Next are those techniques which make use of goal-based questionnaire or a decision tree (formed based on user inputs) to recommend a pattern\cite{Palma2012}\cite{Issaoui2015}. The goal-based techniques seem to offer more accurate results than the previous techniques as the filtering is driven by the user's (architect) choice itself. Techniques like\cite{Issaoui2015} use semantic context, but they need a basic design to start with. The designer is involved in the whole selection process and they are dependent on user's awareness and understanding of the architectural patterns. The closet approach to our work (only in terms of suggesting architectural patterns based on requirements) is \cite{alebrahim2015towards}. 

The proposed system (APR) does not require an architect to start with an initial architecture; APR works directly from the requirements. One major differentiating factor for APR is that while most existing systems focus on design patterns at program level, APR focuses on architectural level patterns. Most existing systems would require changes in their approach in order to produce architectural patterns recommendations. On the other hand, APR is designed to be extensible. To recommend program level patterns APR requires only addition of program level patterns' information in to its knowledge base. In our experiments we have found that the recommendations are as good as the manual approach and it requires minimum effort from a software architect using APR. 

\section{Design of APR}\label{mymethod}
\begin{figure}
\centering
\includegraphics[scale=0.26]{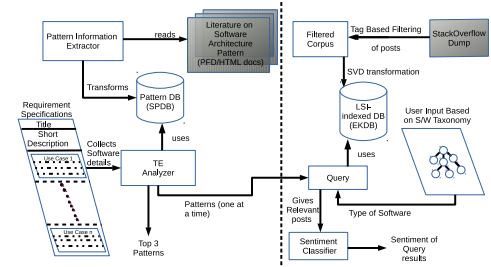}
\caption{APR - The Logical Structure}
\label{overallfig}
\end{figure}

Having access to credible information relevant for making recommendations in a domain of interest, is at the heart of a recommender system. Design of our system relies on established knowledge available in the domain of software architecture design. Two primary sources of such knowledge are: \textbf{(a)} standard text books on software architecture design and \textbf{(b)} professional developer/architect support forums such as Stackoverflow. Major elements comprising APR are as follows:
\vspace*{-2mm}
\begin{enumerate}
\item Architectural Patterns Knowledge-base
\vspace*{-2mm}
\item Textual Entailment Analyzer
\vspace*{-2mm}
\item User Input Template
\vspace*{-2mm}
\item Sentiment Classifier
\end{enumerate}
\vspace*{-3mm}
Details of each of these elements are discussed in the subsections below. The logical structure of our system is shown in Figure-\ref{overallfig} which depicts these four main elements and how they fit in the whole structure.
\vspace*{-3mm}
\subsection{Main Elements of APR}
\label{apr_elements}
\vspace*{-3mm}
\subsubsection{Architectural Patterns Knowledge-base}
\label{apkb}
APR makes use of two different databases that we create - \emph{Standard Patterns Database} and \emph{Experiential Knowledge Database}. \textbf{(a) Standard Pattern Database (SPDB)}: Though different texts follow somewhat varied terminologies to describe and discuss architectural patterns, one way of categorizing architectural pattern is based on the domain of software. For example, distributed system, embedded system, etc. We consider the characteristics of architectural patterns described in Pattern Oriented Software Architecture (POSA) text book\cite{Buschmann1996}, as the basic features on which architectural patterns can be differentiated. Architecture patterns descriptions are also scraped from MSDN\cite{MSDN} as it also describes the patterns in a form similar to POSA\cite{Buschmann1996}. Although, we use two sources to populate SPDB, APR can ingest architectural pattern information from various other sources easily. Internal recommendation generation logic of APR is agnostic to the source of patterns information. The fields of SPDB are depicted in Table-\ref{tab:t2}.
\textbf{(b) Experiential Knowledge Database (EKDB)}: Sheer volume of traffic and the content\footnote{As of September 2016, stackoverflow.com is ranked within top 50 sites globally by Alexa (see http://www.alexa.com/siteinfo/stackoverflow.com); and there are more than 12M questions and 20M answers (see https://data.stackexchange.com/)} available on stackoverflow.com is a strong indication that such developer/architect support forums represent the most updated understanding of a technology by the users of that technology. Particularly, in recent years Stackoverflow\texttrademark  \hspace{1mm} has become a preferred platform for providing developer support by technology vendors. Such forums have questions which are faced by practitioners on day-to-day basis and are answered by those who had faced similar problems. Whether content of the discussion is useful and acceptable solution for the problem, is provided as meta data on most of the forums. We leverage Stackoverflow\texttrademark  \hspace{1mm} data to populate EKDB. Basically, EKDB is a Latent Semantic Indexing (LSI)- indexed database which is discussed in more detail in Section-\ref{sentimet}.  
\begin{table}[!htb]
	\tiny
	 \begin{subtable}[b]{.45\linewidth}
      \centering
        \begin{tabular}{|l|} \hline
            Use Case - ID (ID)\\ \hline
            Name (N)\\ \hline
            Objective (Obj) \\ \hline
            Actors (Act) \\ \hline
            Pre-Conditions (PreCon) \\ \hline
            Post-Conditions (PostCon) \\ \hline
            Constraints (Cst) \\ \hline
            Normal Flow (Flo) \\ \hline
            Importance Score (IS) \\ \hline
        \end{tabular}
        \caption{Sections of Use Case Template}
        \label{tab:t1}
       \end{subtable}
      \begin{subtable}[b]{.45\linewidth}
      \centering
        \begin{tabular}{|l|} \hline
            Pattern Name (PN) \\ \hline
            Basic Definition (BD) \\ \hline
            Context (Ctx) \\ \hline
            Forces (F) \\ \hline
            Solution (Sol) \\ \hline
            Consequences (Conq) \\ \hline
            Variants (Var) \\ \hline
            Known Applications (KA) \\ \hline                       
        \end{tabular}
       \caption{Features of an Architectural Pattern}
       \label{tab:t2}
      \end{subtable}
    \caption{Considered Requirement and Pattern Features}
    \label{two_figs}
\end{table}
\normalsize
\vspace*{-3mm}
\subsubsection{Textual Entailment Analyzer}
To search the standard texts stored in SPDB for a match against user requirements, we use textual entailment recognition techniques. Textual entailment itself makes use of Meaning Representation (MR)\cite{Griffiths2007, Schubert2015}. The MR of a specific text should be all-inclusive and advanced enough to handle rich natural language like English in order to be able to represent English sentences in the form of logical representation (like FOL, DRT, OWL, $\lambda$-calculus based logic etc). We make use of a textual entailment recognizer tool -EOP\cite{Magnini2014} for this purpose. The tool gives a choice between classification based entailment, transformation based entailment and edit distance based. We prefer this tool instead of others as we can use different approaches of entailment, change various parameters and select the best suited one for our domain.  
\vspace*{-3mm}
\subsubsection{User Input Template}
A user of APR is expected to supply the following information in order to get recommendation about suitable architectural patterns that can be used to develop the software application: (i) Short description of the software. (ii) Detailed Description of the software to be developed. (iii) Use cases (in a pre-defined structure) in English to hold the description of various subsystems. (iv) Finally, user has to select the type of software from a drop-down list driven by a taxonomy of software types.  
\vspace*{-3mm}
\subsubsection{Sentiment classifier}
We use the discussion forum content stored in EKDB to determine the sentiment about a recommendation. Such sentiment analysis is used to boost the confidence of recommendations obtained from textual entailment analyzer. For example, suppose APR recommends patterns $p_{1}, p_{2}$ and $p_{3}$ in that order of preference for the architecture of an application $A$. Then the sentiment about the propositions \emph{``use $p_i$ for $A$''} where $i\in(1,2,3)$, is estimated by the classifier.
\vspace*{-3mm}
\begin{algorithm}
\DontPrintSemicolon
\Indentp{0.2em}
\KwIn{$SD$, $DD$, $All\_Obj$, $All\_Act$, $All\_Cst$, $All\_PreCon$, $All\_PostCon$, $NFR$}
\KwOut{Patterns with top $3$ Confidence ($C[p]$), values \\ are returned}
\tcp{Initialize Confidence Array}
$C[p] \gets 0$, where p is index of a pattern\;
\For {each record, r in SPDB}{   
 $C[p] \gets textEntail(DD, r.BD)$\;
 $C[p] \gets C[p]+textEntail(SD, r.KA)$\;
 $C[p] \gets C[p]+textEntail(NFR, r.F)$\;
 $C[p] \gets C[p]+recogEntail(All\_Obj, r.F)$\;
 $C[p] \gets C[p]+recogEntail(All\_Act, r.Sol)$\;
 $C[p] \gets C[p]+recogEntail(All\_Cst, r.F)$\;
 $C[p] \gets C[p]+recogEntail(All\_PreCon, r.Ctx)$\;
 $C[p] \gets C[p]+recogEntail(All\_PostCon, r.Conq)$\;
 }
\caption{Requirement-to-Pattern}
\label{req-pat-algo}
\end{algorithm}
\subsection{Requirements-to-Pattern Textual Analysis}\label{TEsec}
Plain textual matching between user requirement and pattern description do not yield good results. The poor results are due to the way user writes requirement specifications, which might be very different from how a pattern's context is described in literature, even though both might be describing a similar scenario. Keeping this in view we considered user requirements and pattern description to be $text \leftrightarrow hypothesis$ pair when determining textual entailment between them. The algorithm is discussed in detail as follows.

\textbf{Description of inputs to Requirements-to-Pattern algorithm (Algorithm-\ref{req-pat-algo})}- The input arguments expected by this algorithm are as follows:
\vspace*{-3mm}
\begin{enumerate}
\item \textit{Short Description, $SD$}: It describes the software to be developed in maximum $25$ words.
\vspace*{-2mm}
\item \textit{Detailed Description, $D$}: We acquire detailed description as it will provide the user's requirement in one single paragraph without any template. Considering descriptions of a wide range of software applications (both commercial and open source) we selected $500$ words as the limit on this description's length.
\vspace*{-2mm}
\item \textit{Use Cases, $UC$}: The systematic requirements are gathered in the use case structure depicted in Table-\ref{tab:t1}. The requirements are considered in the form of architecturally relevant use cases\cite{Malan2001}. We take input in terms of use cases because they are concise way of writing functional scenarios in requirement phase\cite{Potts1994}. Also, use cases can be defined at different levels and different perspectives\cite{Kettenis2007}. Primarily, as our input we consider the use cases that are at ``summary level" or ``user goal level"\cite{Kettenis2007}. As implied by \cite{Potts1994,Kettenis2007}, we believe that utilizing the use cases in this manner is an effective way to suggest a system wide architectural pattern. 
Architecturally relevant use cases affect the overall system, and they generally have higher importance. In Table-\ref{tab:t1} the last field of the use case template is ``Importance Score". This field signifies the importance given to each use case by the user. The accepted value of ``Importance Score"  field lies between $0$ and $1$. Here $0$ implies least significance and $1$ implies the most. If not mentioned, the default value is taken to be $1$. Rest all fields of the use case template are self-explanatory. For a given software application the user may provide use cases within a range of $1-20$. If the number of use cases is more than 1 (which will be the case usually), we call Algorithm-\ref{init-algo} to create sets of tuples consisting of data from similar fields of use cases with their respective importance score. The set of tuples returned by this function is given as input to Algorithm-\ref{req-pat-algo}.
\vspace*{-2mm}
\item \textit{Non-Functional Requirements, $NFR$}: A set of non-functional requirements that must be satisfied by the desired software application is taken from the user. Conflicts among supplied NFRs are checked based on the classification provided in\cite{Mairiza2010}. If none of the supplied NFRs contradict they are considered in the input NFR set else user is asked to give priority to the contradicting NFRs. The one with the highest priority is kept. 
\end{enumerate}
\vspace*{-3mm}
In addition to these inputs, Algorithm-\ref{req-pat-algo} makes use of two procedures- (i) $textEntail(string_{1}, string_{2})$. It determines textual entailment between $string_{1}$ and $string_{2}$ using EOP\cite{Magnini2014}. Various internal parameters in EOP were determined after repeated training/testing iterations on our data. (ii) $recogEntail(set\_of\_tuple, Pattern\_Attr)$. This procedure determines textual entailment between a set of tuples of the form $\langle string, importance_{string}\rangle$. $textEntail(string_{1}, string_{2})$ is called by it wherever necessary. The procedure is depicted in Algorithm-\ref{calmap-algo}. A mapping between pattern features and use case fields have been defined in Table-\ref{tab:mapping}. When finding patterns to recommend using textual entailment recognition, we use this mapping to determine a pattern's features that correspond to a field in the input use case.
\vspace*{-1mm}
\begin{algorithm}
\DontPrintSemicolon
\Indentp{0.2em}
\KwIn{$UC$}
\KwOut{Sets containing aggregated use case fields of respective type.}
\tcp{Initializing sets}
$All\_Obj \gets \{\}$, $All\_Act \gets \{\}$, $All\_Cst \gets \{\}$, \\ $All\_PreCon \gets \{\}$, $All\_PostCon \gets \{\}$\;
\For{each use case, $uc \in UC$}{
  $All\_Obj = \bigcup \langle uc.Obj, uc.IS \rangle$\; 
  $All\_Act = \bigcup \langle uc.Act, uc.IS \rangle$\; 
  $All\_Cst = \bigcup \langle uc.Cst, uc.IS \rangle$\;
  $All\_PreCon = \bigcup \langle uc.PreCon, uc.IS \rangle $\;
  $All\_PostCon = \bigcup \langle uc.PostCon, uc.IS \rangle $\;
  }
\Return $All\_Obj, All\_Act, All\_Cst, All\_PreCon, All\_PostCon$
\caption{Creating Sets of Use case's Fields }
\label{init-algo}
\end{algorithm}
\vspace*{-7mm}
\setlength{\intextsep}{1\baselineskip}
\begin{algorithm}
\DontPrintSemicolon
\Indentp{0.2em}
\KwIn{$SetOfUseCases$, $Pattern\_Attr$}
\KwOut{$Entail\_value$}
\tcp{Initialize confidence Value}
$CV \gets 0$\; 
\For{each $tuple, t, in~SetOfUseCases$}{
    $x \gets 0$\;
    \tcp{t[0] contains usecase field}
 	$x \gets textEntail(t[0], Pattern\_Attr)$\;
 	\tcp{t[1] contains importance score for t[0]}
 	$x \gets x * (t[1])$\;
 	$CV = CV + x$\;
 	}
\Return $CV$
\caption{recogEntail}
\label{calmap-algo}
\end{algorithm} 

\vspace*{-1mm}
\begin{table}[h]
\centering
\begin{tabular}{|p{4cm}|l|}
\hline
\textbf{Requirements specs/Usecase section} & \textbf{SPDB column} \\ \hline
Detailed definition (DD) & Basic definition (BD) \\ \hline
Short description (SD) & Known applications (KA) \\ \hline
Objectives (Obj) & Forces (F) \\ \hline
Post-conditions (PostCon) & Consequences (Conq) \\ \hline
Constraints (Cst) & Forces (F) \\ \hline
Pre-conditions (PreCon) & Context (Ctx) \\ \hline
Actors (Act) & Solution (Sol) \\ \hline
Flow (Flo) & Solution (Sol) \\ \hline
Non-functional requirements (NFR) & Forces (F) \\ \hline
\end{tabular}
\caption{Requirements fields to pattern feature mappings}
\label{tab:mapping}
\end{table}
\vspace*{-6mm}

\vspace*{-2.5mm}
\subsection{Sentiment Analysis of APR Recommendations}\label{sentimet}
In Section-\ref{apkb}(a) we have discussed the motivation for leveraging crowd-sourced knowledge available on developer/architect support forums such as Stackoverflow. As a final step APR calculates the sentiment/opinion about the recommendations it generates. In order to calculate sentiment scores the APR relies on EKDB database. EKDB is created from the dump\footnote{Available here https://archive.org/download/stackexchange} of Stackoverflow posts. Specifically, a subset of the posts available in the dump are filtered based on their relevance for architectural pattern selection for different kinds of software. We use tags data available in each post to decide about its relevance\footnote{This may leave out some posts which are untagged but are relevant. Discussed in Section-\ref{ttv}}. We then index the filtered posts using Latent Semantic Indexing (LSI)\cite{Wiemer-Hastings2004}.

The steps involved in deriving sentiment score for recommendations generated by APR are described as follows:
\vspace*{-3mm}
\begin{enumerate} 
\item Let the architectural patterns recommended by APR be an ordered list $P$. For discussion sake, say,  $P = \{p_{1}, p_{2}, p_{3}\}$. 
\vspace*{-2mm}
\item User selects the ``type" (let's call it $T$) of the desired software application from a hierarchical list driven by the software taxonomy provided in\cite{Forward2008}. Selecting type of the software application provides a suitable context for querying EKDB.
\vspace*{-2mm}
\item For each pattern recommendation $p_i$ we synthesise a query $q_i$ as follows: $q_i = p_i$ + ``for'' + $T$, where + means concatenation. For example, if $p_{1} = $ \texttt{MVC} and $T = $ \texttt{Web Based Application} then $q_{1} = $ \texttt{MVC} \texttt{for Web Based Application}.
\vspace*{-2mm}
\item Result set ($R$) obtained by running each query $q_i$, on EKDB is iterated over, and for each record we calculate its sentiment using English words provided by \cite{Nielsen2011}\footnote{We have further enhanced this list with terms from our domain like - handles, overkill, separation of logic, natural fit and few others.}. The cumulative sentiment is determined by simple addition\footnote{Sentiment aggregation will be improved in future enhancements to APR.} of individual values for each record.
\vspace*{-2mm}
\end{enumerate}

\vspace*{-1mm}
The sentiment scale used is: \texttt{strongly positive, positive, slightly positive, neutral, slightly negative, negative, strongly negative}. If EKDB does not contain any post regarding a particular query we consider the sentiment to be neutral.

\section{Experiments Analysis}\label{Analy}

We ran APR for $30$ different applications covering different categories of software. To ensure variety we chose between 2 to 5 applications in each category of software applications. For example, we framed specifications for a `simple complier', `expert system', `stock-ticker mobile app' etc. For each application, in addition to its description we came up with architecturally relevant use cases required to implement its functionality. We analyzed the performance of APR system in terms of ``ground truths". Here, by ground truths we mean architectural pattern used by existing software systems of similar nature whose architecture details were well known. We compared our results with the architectural pattern actually used by similar software. Results obtained from APR for a subset of such 30 applications are depicted in Table-\ref{AnalysisTable}.

We observed that about $85\%$ of the $1^{st}$ ranked recommendations of APR matched with the architectural pattern used in reference application of similar nature (i.e. ``ground-truth"). In the remaining $15\%$ of the recommendations, the pattern was not very different from what was being used in the ``ground-truth" case. For example, when the application type was ``mini compiler" the correct answer would have been `Pipe-and-Filter' but APR recommended `Layers'(as $1^{st}$ choice, the $2^{nd}$ choice was `Pipe-and-Filter') but did not digress as far as `MVC' or `PAC'. Also, in these $15\%$ cases the $2^{nd}$ or $3^{rd}$ ranked pattern is the correct suggestion. First column of Table-\ref{tab:Result-Analy} depicts the exact number of application for which APR gave expected output and at which rank. We can see that for two software applications the expected pattern was the $3^{rd}$ ranked, for three it was $2^{nd}$ ranked and for twenty-five applications the expected pattern was given as the top ranked output. The second and third column of Table-\ref{tab:Result-Analy} show what was the popular sentiment of the variously ranked pattern. For example, in $row1, column2$ we can observe that in $18$ cases the top ranked pattern had garnered positive sentiment in Stackoverflow posts.  
\vspace*{-0.5mm}
\begin{table}[h]
\centering
\begin{tabular}{p{1cm}|p{2cm}|p{2cm}|p{2cm}} \hline
             & \textbf{Expected Output}&\textbf{Positive Sentiment}&\textbf{Negative Sentiment} \\ \hline \hline
$1^{st} rank$& 25 & 18 & 3\\ \hline
$2^{nd} rank$& 3  & 14 & 5\\ \hline
$3^{rd} rank$& 2  & 6  & 0\\ \hline
\end{tabular}
\caption{APR - Output Analysis}
\label{tab:Result-Analy}
\end{table}

\vspace*{-3mm}

\newcommand{\ra}[1]{\renewcommand{\arraystretch}{#1}}
\begin{table*}[!t]
\tiny
\centering
\ra{0.8}
\begin{tabular}{@{}p{2cm}p{2cm}p{8cm}p{0.5cm}|p{1cm}p{0.3cm}p{2cm}@{}}
\toprule
\multicolumn{4}{c}{Requirements} & \multicolumn{2}{c}{Patterns}  
\\
\cmidrule{1-4} \cmidrule{5-7} 
Title & Type &Description& No. of Use Case & Pattern & Rank & Sentiment \\ \midrule
Content Management System for a University&Business oriented software for University&The platform should help users to build and maintain web pages of a particular university. Pre-approved users should be able to publish on-line without programming. It should provide an integrated workflow. Also, users should have different levels of permissions. The pages to be published should be verified by particular users. It should be driven by data to have uniform changes across all pages. It should also be able to manage record of day to day administrative decisions and send reminder to concerned person if some information needs to changed.&13& MVC&1&Strongly Positive\\
&&&& PAC&2& Neutral\\
&&&& Microkernel&3& Positive \\ \midrule
To provide a tool which handles compatibility issues while execution of a software program' &Design and Engineering Software for Development Environment& This tool should help the user to run a source code written in a particular environment to run on other one based on the specifications given. In other words it should be able to create the environment specified by the user in a virtual environment.&5&Microkernel&1&Positive \\
&&&& Layers&2& Neutral\\
&&&& Reflection&3& Neutral \\ \midrule
To provide shell like capabilities in a limited manner&System software especially emulator&The software should be a lightweight working shell on Unix Environment. It should include all the basic functionalities of a Unix shell. It should be usable for academic purposes. It should be easily extend-able/modifiable such that the students can learn practically from it&8&Pipe-and-Filter&1&Strongly Positive \\
&&&& Microkernel&2&Slightly Positive \\
&&&& Layered&3&Neutral \\
\bottomrule
\end{tabular}
\caption{Recommendations and Sentiment Based on Few Scenarios}
\label{AnalysisTable}
\end{table*}

\normalsize
\subsection{Few Results}
Table-\ref{AnalysisTable} depicts few of the scenarios out of $30$.  The first four columns depict the user input. For want of space here, we simply provide the number of use cases that were taken as input. Outputs depicted in $4^{th}$, $5^{th}$ and $6^{th}$ column illustrate APR's recommendation along with sentiment information. A value ``neutral" in last column may also be due to non-existence of any data regarding that particular scenario. For verification purpose we consider Joomla\cite{Joom}, Vagrant\cite{mitchellh} and Powershell\cite{Microsoft} software and their architectural patterns. Main reason for choosing these particular software applications is the availability of architectural documentation for them. Consider the case where the user is seeking architectural pattern recommendations for designing a \texttt{Web based Application}. This case is shown in row No. 1 of Table-\ref{AnalysisTable}. 
$1^{st}$ row in this table depicts the results for a case where the user is seeking architectural pattern recommendations from APR about \texttt{Content Management System (CMS)}. APR gives MVC pattern as the top most recommendation. We looked at popular CMS software such as Joomla, Drupal, WordPress etc. All of them are web applications which use MVC or a variant of MVC as the principal architectural pattern, thus validating the correctness of APR's recommendation. The sentiment analysis results also show that most people strongly feel MVC to be the correct choice for a content management system. Similarly, $2^{nd}$ and $3^{rd}$ row correspond to a software similar to Vagrant and Powershell respectively. Vagrant uses something like Layers pattern whereas Powershell is based on Pipe and Filter pattern which is what is also recommendebad by APR as $2^{nd}$ and $1^{st}$ suggestion.  
\vspace*{-3mm}
\subsection{Threats to Validity}\label{ttv}
Although APR performs well, there are few scenarios where APR has some limitations. One of our assumption while generating recommendations in APR is that the software to be developed will use/require a single architectural pattern. However, in some complex software applications this assumption may not hold good. As such, the recommendations may not be accurate in those cases. To address this gap we are working towards incorporating hierarchical use cases as input to APR. In our current experiments the APR built the patterns knowledge base (SPDB) from mainly two (although seminal texts) sources, which may be seen as a limitation. However, it is relatively easy to incorporate patterns informations/knowledge from other authentic sources as well. Next, the posts form Stackoverflow which are considered suitable to be added in APR knowledge base (EKDB) are those which have relevant tags on them. Since all posts on stakoverflow are not properly tagged so we might be missing few posts which do not have the relevant tags but are still admissible. Though this limitation can be addressed by changing the query so that it considers all posts, it will become much slower. Another point which may be seen as a potential limitation of APR is the simple addition based method being currently used for aggregating sentiment scores. In this context it may be worth noting that a pattern which is used heavily or is more popular may have more strong emotions (-ve or +ve) than the one which is used less. This may lead to skewing of recommendations as we observed in our results. We are working towards weighing emotions according to the popularity of patterns. Finally, our semantic classifier is generic and not specific\footnote{Although we had added good number of words from software engineering domain as discussed in Section-\ref{sentimet}} to software engineering domain. Due to this we may have ignored few words and marked them in wrong manner.

\section{Conclusion and Future Work}\label{confut}
We have proposed a recommender system, APR, to assist a software architect in the task of selecting a suitable architectural pattern for a given set of software requirements. Our objective is to minimize the architect's effort by semi-automating the manual process involved in identifying candidate architectures that can fulfil a given set of requirements. To achieve superior accuracy of searching and recommending an architectural pattern that meets a given set of software requirements APR employs textual entailment recognition techniques. 
APR also calculates sentiment (+ve to -ve scale) about each of the recommendations by considering the relevant discussions on stackoverflow. 
Such sentiment classification further helps the software architect in determining whether or not to use a particular pattern in a specific scenario. Our evaluation experiments with APR show that suggestion list given by APR are similar to what an experienced architect would have selected in a similar situation while manually sifting through books and information from the Web. That said, we have also noted the limitations of APR in certain areas. For example, a better sentiment classifier which is suited for the domain of software architecture is required to get more accurate results when calculating sentiment scores for the recommendations. In future we would like to address those limitations.


%
\bibliographystyle{abbrv}
\bibliography{sigproc}  
%
%

\end{document}